# Measures of Helicity and Chirality of Optical Vortex Beams

Kayn A. Forbes and Garth A. Jones

*School of Chemistry, University of East Anglia, Norwich, Norfolk, NR4 7TJ, United Kingdom*

k.forbes@uea.ac.uk

**Abstract** Analytical forms of the optical helicity and optical chirality of monochromatic Laguerre-Gaussian optical vortex beams are derived up to second order in the paraxial parameter $kw_0$. We show that input linearly polarised optical vortices which possess no optical chirality, helicity or spin densities can acquire them at the focal plane for values of a beam waist $w_0 \approx \lambda$ via an OAM-SAM conversion which is manifest through longitudinal (with respect to the direction of propagation) fields. We place the results into context with respect to the intrinsic and extrinsic nature of SAM and OAM, respectively; the continuity equation which relates the densities of helicity and spin; and the newly coined term 'Kelvin's chirality' which describes the extrinsic, geometrical chirality of structured laser beams. Finally, we compare our work (which agrees with previous studies) to the recent article Köksal, *et al.* Optics Communications **490,** 126907 (2021) which shows conflicting results, highlighting the importance of including all relevant terms to a given order in the paraxial parameter.

1. Introduction

A stationary object is chiral if it cannot be superimposed onto its mirror image. Material chirality is a consequence of the geometric position of atoms in molecules and nanostructures. In order to discriminate the chirality of an object in an enantiospecific manner, the probe must also exhibit a chiral structure. Chiroptical and optical activity spectroscopies – using differential light-matter interactions – are widely used techniques to study the structures and functionalities of chiral molecules, biomolecules, metamaterials and nanostructures, as well as achiral objects such as atoms [1–12]. The observables in these spectroscopies are either time-even pseudo scalars or time-odd axial vectors; the former often referred to as true chirality on the account that these effects require chiral material structures, the latter false chirality as they can be supported by achiral media such as atoms as in magnetic circular dichroism [13].

Light most commonly exhibits chirality through circular polarisation (CPL), a local property, where the electric and magnetic field vectors trace out a helical pattern on propagation. The handedness of CPL is denoted by $\sigma = \pm 1$, often referred to as the helicity, with the positive value corresponding to left-handed CPL, and the negative to right-handed CPL. Less well-known is the fact that light can also possess an additional type of chirality due to its spatial structure, a global property; examples include those stemming from polarisation structure or phase structure [14]. This geometrical chirality of structured light has recently been termed 'Kelvin's chirality' [15]. In this work we are specifically interested in the chirality of scalar beams that possess a phase structure of the form $e^{i\ell\phi}$ where $\phi$ is the azimuthal angle: these modes are often referred to as optical vortices. Analogous to $\sigma$ for CPL, the sign of the pseudo scalar topological charge $\ell \in \mathbb{Z}$ is what determines the handedness of an optical vortex: $\ell > 1$ the vortex is left handed; $\ell < 1$ it is right-handed.

Optical vortices have found widespread application since the early 90s – the reader is referred to a few of the many review articles [16–18] – and recently their chirality has been directly utilised in chiroptical spectroscopies [19] and chiral nanostructure fabrication [20].

The conserved pseudo-scalar quantities known as optical helicity and optical chirality (we use lowercase for the local conserved quantity optical chirality of beams; we refer to the geometric chirality of beams as Kelvin's chirality) of electromagnetic fields have found distinct theoretical application in recent years in chiroptical interactions. The optical chirality has been widely studied since Tang and Cohen highlighted its relevance in light-matter interactions [21–30]. For overviews see Refs [31,32]. Closely related is the optical helicity, which for monochromatic fields is proportional to the optical chirality [33].

In this work we undertake a systematic study of the optical chirality and helicity of optical vortex beams. We begin in Section 2 with a brief recap of the general definitions of optical chirality and helicity in free space, followed by giving the necessary quantum electromagnetic mode operators to undertake the calculation of these conserved quantities for Laguerre-Gaussian optical vortices in Section 3. We then derive the analytical formula up to second-order in the paraxial parameter for both circularly polarised and linearly polarised modes in Sections 4 and 5, highlighting the particularly interesting case of an optical helicity (Chirality) even for linearly polarised modes, as well as the spin-orbit interactions that occur in circularly polarised modes. Furthermore, we comment on a recent study which gives conflicting results to those derived here. In Section 6 we look at the spin angular momentum density as it is directly linked to the optical helicity through a continuity equation. We conclude with a discussion of the results in Section 7.

## 2. Optical helicity and optical chirality in free space

The most general definitions of the optical helicity $\mathcal{H}$ and the optical chirality $X$ in free space are usually given in terms of the electromagnetic potentials $A$ and $C$ and electromagnetic fields $E$ and $B$ [34]:

$$\mathcal{H} = \int d^3 r \frac{\varepsilon_0 c}{2} (A \cdot B - C \cdot E), \tag{1}$$

$$X = \int d^3 r \frac{\varepsilon_0}{2} (E \cdot \nabla \times E + c^2 B \cdot \nabla \times B), \tag{2}$$

where $B = \nabla \times A$ and $E = -\nabla \times C$. Both integrated quantities (1) and (2) are gauge-invariant and Lorentz-covariant, however the integrand of (1) is not gauge-invariant. Throughout this work we deal with gauge invariant integrands at the expense of losing Lorentz-covariance by working within the Coulomb gauge $\nabla \cdot A = 0$ and $\nabla \cdot C = 0$. This is an easily justifiable action in the field of optics. The integrands in (1) and (2) are known respectively as the optical helicity density $h$ and optical chirality density $\chi$. The introduction of the potential $C$ ensures that (1) retains its form under duplex transformation [35,36], i.e. the expressions (1) and (2) are dual symmetric. The necessity of these quantities being dual symmetric is most obvious when one realises, they are conserved for the free field, i.e. $\partial_t \mathcal{H}, \partial_t X = 0$. Thorough discussions on some of the more fundamental and subtle aspects of the form and properties of (1) and (2) can be found in references [31,32,37].

We will be using a quantum electrodynamic (QED) formulation in the Coulomb gauge [38] throughout this work, and so the fields and potentials are the microscopic operator versions of those in (1) and (2). Furthermore, compared to numerous studies which utilise natural units $\varepsilon_0 = \mu_0 = c = 1$ we explicitly use S.I. units. The optical helicity $\mathcal{H}$ may be given in QED as

$$\mathcal{H} = \int d^3 r \frac{\varepsilon_0 c}{2} \left( \boldsymbol{a}^\perp \cdot \boldsymbol{b} - \boldsymbol{c}^\perp \cdot \boldsymbol{e}^\perp \right), \tag{3}$$

where $\boldsymbol{a}^\perp$ ($\boldsymbol{c}^\perp$) is a vector potential operator; $\boldsymbol{e}^\perp$ is the transverse electric field operator; and $\boldsymbol{b}$ is the magnetic field operator. The superscript $\perp$ denotes transversality of the field with respect to the Poynting vector in this gauge (not necessarily the direction of propagation); $\boldsymbol{b}$ is purely transverse irrespective of gauge [38]. The integrand of (3) is the helicity density $h$. In the QED theory utilised here we note that (3) does not require the so-called electric helicity contribution $\boldsymbol{C} \cdot \boldsymbol{E}$ of (1) necessary in the classical theory to satisfy the conservation law $\partial_t \mathcal{H} = 0$. However, it still requires the introduction of the $\boldsymbol{C} \cdot \boldsymbol{E}$ in order to satisfy dual symmetry required in the continuity equation with the spin angular momentum density $\boldsymbol{s}$ [34,36,39]:

$$\dot{h} + \nabla \cdot \frac{\varepsilon_0}{2} \left( \boldsymbol{e}^\perp \times \boldsymbol{a}^\perp + \boldsymbol{b} \times \boldsymbol{c}^\perp \right) = 0, \tag{4}$$

where the quantity $\frac{\varepsilon_0}{2} \left( \boldsymbol{e}^\perp \times \boldsymbol{a}^\perp + \boldsymbol{b} \times \boldsymbol{c}^\perp \right)$ is the correct dual symmetric spin density $\boldsymbol{s}$ of the free field.

The optical chirality $X$ in QED is defined as

$$X = \int d^3 r \frac{\varepsilon_0}{2} \left( \boldsymbol{e}^\perp \cdot \nabla \times \boldsymbol{e}^\perp + c^2 \boldsymbol{b} \cdot \nabla \times \boldsymbol{b} \right). \tag{5}$$

The integrand of (5) is the optical chirality density $\chi$. For monochromatic fields the optical helicity (3) and optical chirality (5) are proportional, with the correct proportionality constant being $\omega k = c k^2 = \omega^2 / c$ as opposed to the routinely reported factor $\omega^2$ due to the more common use of natural units. In this work we study the optical helicity density explicitly as it is a more fundamental quantity, in the knowledge that the optical chirality gives an analogous result for the system we are studying. Due to being in the Coulomb gauge we can re-write (3) in terms of the electromagnetic fields only

$$\mathcal{H} = \varepsilon_0 c \int d^3 \boldsymbol{r} \left( -\int \boldsymbol{e}^\perp dt \right) \cdot \boldsymbol{b}, \tag{6}$$

where we have used the fact that $-\dot{\boldsymbol{a}}^\perp = \boldsymbol{e}^\perp$ and $-\dot{\boldsymbol{c}}^\perp = \boldsymbol{b}$; both (6) and the spatial integrand of (6) are clearly gauge-invariant. It is highly interesting to note that the expressions (3) and (6) hold in either the standard (asymmetric, electric biased) or dual-symmetric theories, all other conserved quantities, like energy density or angular momentum density, take on different forms in either the asymmetric or and dual symmetric theories [37].

### 3. Electromagnetic field mode operators

In order to evaluate the quantities $h$ and $\chi$ from the previous section we require the form of the electromagnetic field mode expansion operators $\boldsymbol{e}^\perp$ and $\boldsymbol{b}$. Numerous studies in optics and light-matter interactions assume $\boldsymbol{e}^\perp$ and $\boldsymbol{b}$ to be purely transverse to the direction of propagation, which itself exactly coincides with the Poynting vector. This plane-wave description of radiation hides a plethora of novel properties of radiation fields in real world optical and nano optical experiments involving focused or spatially-confined light [40]. In order to strictly satisfy Maxwell's equations no physically realisable radiation field is completely transverse. It was Lax *et al.* [41] who first undertook a systematic study of this issue with respect to paraxial light fields, highlighting how to generate the higher-order fields necessary for a completely correct description of a radiation field. In general, a radiation field consists of a usually dominant zeroth-order transverse component $\boldsymbol{T}_0^{e/b}$ (the term most resembling a plane-wave description), plus higher, odd-integer-order $\boldsymbol{L}_{2n+1}^{e/b}$ longitudinal components and even-integer transverse components $\boldsymbol{T}_{2n}^{e/b}$. For example, first-order $\boldsymbol{L}_1^{e/b}$ and third-order $\boldsymbol{L}_3^{e/b}$ longitudinal fields; second-order transverse fields $\boldsymbol{T}_2^{e/b}$, and so on. The reader is referred to the following further references which discuss this topic [42–45]. Each higher-order field component is weighted by the so-called paraxial factor $n(kw_0)^{-1}$, where $k$ is the wave number and $w_0$ is the beam waist at the focal plane $z=0$, and as such the magnitude of higher-order terms are directly related to the degree of focusing (or confinement).

In this work we specifically concentrate on the Laguerre-Gaussian (LG) optical vortex mode, a solution to the paraxial wave equation in cylindrical coordinates. These are the most widely utilised type of optical vortex beam in experiments. The LG mode is characterised by four parameters $(k, \eta, \ell, p)$: the wave number $k$; polarisation $\eta$; topological charge $\ell$; and radial index $p$ with $(p+1)$ indicating the number of radial nodes. The zeroth-order electric field $\mathrm{T}_0^e$ mode expansion operator $\boldsymbol{e}^\perp(\boldsymbol{r})$ for a LG mode in the long Rayleigh range $z_R \gg z$, where $2z_R = kw_0^2$, is given by [46]:

$$\boldsymbol{e}^\perp(\boldsymbol{r}) = i \sum_{k,\eta,\ell,p} \left( \frac{\hbar c k}{2\varepsilon_0 A_{\ell,p}^2 V} \right)^{1/2} \left[ \boldsymbol{e}^{(\eta)}(k\hat{z}) f_{|\ell|,p}(r) a_{|\ell|,p}^{(\eta)}(k\hat{z}) \mathrm{e}^{i(kz+\ell\phi)} - \mathrm{H.c} \right], \tag{7}$$

where $\boldsymbol{e}^{(\eta)}(k\hat{z})$ is the polarisation unit vector for a z-propagating, $\eta$ polarised, photon; $a_{|\ell|,p}^{(\eta)}(k\hat{z})$ is the annihilation operator; $\mathrm{e}^{i(kz+\ell\phi)}$ is the phase factor, the azimuthal dependent part being responsible for the optical OAM; $f_{|\ell|,p}(r)$ is a radial distribution function (see Appendix A); $V$ is the quantisation volume; $A_{\ell,p}$ is a normalisation constant; and H.c. stands for Hermitian conjugate. The zeroth-order transverse field (7) gives a realistic physical picture provided $(kw_0)^2 \gg 1$ [47].

With the aid of Faraday's and Ampere's laws we can calculate the first-order longitudinal $L_1^e$ and second-order transverse $T_2^e$ contributions to (7), as well as the zeroth and second-order transverse, $T_0^b$ and $T_2^b$, respectively, and first-order longitudinal magnetic fields $L_1^b$. Namely, Faraday's law produces

$$\begin{aligned} \boldsymbol{b}(\boldsymbol{r}) &= -\int (\nabla \times \boldsymbol{e}^{\perp}(\boldsymbol{r})) dt \\ &= -\frac{i}{ck}(\nabla \times \boldsymbol{e}^{\perp}(\boldsymbol{r})) \\ &= i \sum_{k,\eta,\ell,p} \left(\frac{\hbar ck}{2\varepsilon_0 A_{\ell,p}^2 V}\right)^{1/2} \left[-\frac{i}{ck}(\nabla \times \boldsymbol{e}^{(\eta)}(k\hat{\boldsymbol{z}})) f_{|\ell|,p}(r) a_{|\ell|,p}^{(\eta)}(k\hat{\boldsymbol{z}}) e^{i(kz+\ell\phi)} - \text{H.c}\right]. \end{aligned} \quad (8)$$

Then inserting (8) into Ampere's Law

$$\begin{aligned} \boldsymbol{e}^{\perp}(\boldsymbol{r}) &= c^2 \int (\nabla \times \boldsymbol{b}(\boldsymbol{r})) dt = \frac{ci}{k}(\nabla \times \boldsymbol{b}(\boldsymbol{r})) \\ &= i \sum_{k,\eta,\ell,p} \left(\frac{\hbar ck}{2\varepsilon_0 A_{\ell,p}^2 V}\right)^{1/2} \left[\frac{1}{k^2}(\nabla \times \nabla \times \boldsymbol{e}^{(\eta)}(k\hat{\boldsymbol{z}})) f_{|\ell|,p}(r) a_{|\ell|,p}^{(\eta)}(k\hat{\boldsymbol{z}}) e^{i(kz+\ell\phi)} - \text{H.c}\right] \\ &= i \sum_{k,\eta,\ell,p} \left(\frac{\hbar ck}{2\varepsilon_0 A_{\ell,p}^2 V}\right)^{1/2} \left[\frac{1}{k^2}(\nabla(\nabla \cdot \boldsymbol{e}^{(\eta)}(k\hat{\boldsymbol{z}})) - \nabla^2 \boldsymbol{e}^{(\eta)}(k\hat{\boldsymbol{z}})) f_{|\ell|,p}(r) a_{|\ell|,p}^{(\eta)}(k\hat{\boldsymbol{z}}) e^{i(kz+\ell\phi)} - \text{H.c}\right]. \end{aligned} \quad (9)$$

Note we could have also used Gauss's law and $\nabla \cdot \boldsymbol{b} = 0$ to calculate the first-order longitudinal fields, sometimes referred to as the 'transversality conditions of Maxwell's equations'. The explicit forms of the individual contributions $\boldsymbol{T}_{2n}^{e/b}$ and $\boldsymbol{L}_{2n+1}^{e/b}$ for both circular and linear polarisation can be found in Appendix C.

### 4. T$_0$ and L$_1$ contributions to the optical helicity

**4.1**

In calculating $h$ with (8) and (9) there is the possibility of nine distinct terms:

$$h \propto \boldsymbol{e}^{\perp} \cdot \boldsymbol{b} \propto \left(\boldsymbol{T}_0^e + \boldsymbol{L}_1^e + \boldsymbol{T}_2^e\right) \cdot \left(\boldsymbol{T}_0^b + \boldsymbol{L}_1^b + \boldsymbol{T}_2^b\right), \quad (10)$$

however it is clear that any cross-terms between longitudinal and transverse components in the dot product will be zero: $(\hat{x}, \hat{y}) \cdot \hat{z} = 0$ and so there are only five terms that require investigation:

$$h \propto \boldsymbol{T}_0^e \cdot \boldsymbol{T}_0^b + \underbrace{\boldsymbol{T}_0^e \cdot \boldsymbol{T}_2^b + \boldsymbol{T}_2^b \cdot \boldsymbol{T}_0^e + \boldsymbol{L}_1^e \cdot \boldsymbol{L}_1^b}_{(kw_0)^{-2}} + \underbrace{\boldsymbol{T}_2^e \cdot \boldsymbol{T}_2^b}_{(kw_0)^{-4}}. \tag{11}$$

The zeroth-order pure term takes on its maximum value for circular polarisation and is zero for linearly polarised inputs. The interest moving beyond this zeroth-order approximation often made is twofold: Firstly, other terms in (11) are not necessarily zero even for linearly polarised optical vortices; and secondly that spin-orbit-interactions of light (SOI) occur in optical vortices through higher-order components of the field.

In this Section we calculate the helicity density contributions from the pure zeroth-order fields $\boldsymbol{T}_0^e \cdot \boldsymbol{T}_0^b$ and the first-order longitudinal fields $\boldsymbol{L}_1^e \cdot \boldsymbol{L}_1^b$. We give analytical results for the important cases of a circularly polarised input and a linearly polarised input in the *x*-direction but note results are in general polarisation-dependent. In the next Section we calculate the $\boldsymbol{T}_0^e \cdot \boldsymbol{T}_2^b$ and $\boldsymbol{T}_2^b \cdot \boldsymbol{T}_0^e$ contributions; we do not calculate the $\boldsymbol{T}_2^e \cdot \boldsymbol{T}_2^b \propto 1/(kw_0)^4$ term as it significantly smaller than the other terms in general (See Section 7).

*4.2 circular polarisation*

To the level of $\boldsymbol{T}_0^{e/b}$ and $\boldsymbol{L}_1^{e/b}$ the field mode operators take on the following form for a circularly polarised input [48]

$$\boldsymbol{e}^\perp(\boldsymbol{r}) = i \sum_{k,\sigma,\ell,p} \left( \frac{\hbar ck}{2\varepsilon_0 A_{\ell,p}^2 V} \right)^{1/2} \frac{1}{\sqrt{2}} \left[ \left\{ (\hat{\boldsymbol{x}} + i\sigma\hat{\boldsymbol{y}}) + \frac{i}{k}\hat{\boldsymbol{z}} \left( \frac{\partial}{\partial r} - \ell\sigma\frac{1}{r} \right) e^{i\sigma\phi} \right\} \right. \\ \left. \times f_{|\ell|,p}(r) a_{|\ell|,p}^{(\sigma)}(k\hat{\boldsymbol{z}}) e^{i(kz+\ell\phi)} - \text{H.c} \right], \tag{12}$$

and

$$\boldsymbol{b}(\boldsymbol{r}) = i \sum_{k,\sigma,\ell,p} \left( \frac{\hbar k}{2c\varepsilon_0 A_{\ell,p}^2 V} \right)^{1/2} \frac{1}{\sqrt{2}} \left[ \left\{ (\hat{\boldsymbol{y}} - i\sigma\hat{\boldsymbol{x}}) + \frac{1}{k}\hat{\boldsymbol{z}} \left( \sigma\frac{\partial}{\partial r} - \frac{\ell}{r} \right) e^{i\sigma\phi} \right\} \right. \\ \left. \times f_{|\ell|,p}(r) a_{|\ell|,p}^{(\sigma)}(k\hat{\boldsymbol{z}}) e^{i(kz+\ell\phi)} - \text{H.c} \right]. \tag{13}$$

Assuming an input monochromatic field mode of $|n(k,\sigma,\ell,p)\rangle$ photons and inserting (12) and (13) into (6) gives an optical helicity density

$$h = \sum_{\sigma,\ell,p}\left(\frac{n\hbar}{A_{\ell,p}^2 V}\right)\left[\sigma f^2 + \frac{1}{2k^2}\left(\sigma f'^2 - \frac{2\ell}{r}ff' + \frac{\ell^2\sigma}{r^2}f^2\right)\right], \tag{14}$$

where we now drop the dependencies of the factors for notational brevity; the forms of $ff'$ and $ff'$ can be found in the Appendix A (the prime denoting partial differentiation with respect to $r$). We see that (14) has units of angular momentum density as required. The integrated optical helicity $\mathcal{H}$ of (14) gives the expected result of being the difference in number left and right-handed circularly polarised photons.

There are three distinct combinations of $\sigma$ and $\ell$ which lead to differing forms of (14) [49]. When $\ell = 0$ (i.e. a Gaussian beam) we yield optical helicity densities in the focal plane which are purely down to the circular polarisation state $\sigma = \pm 1$, these are plotted in Figure 1. All figures throughout the manuscript correspond to a value of $w_0 = \lambda$ and it must be remembered that the phenomena which arise from longitudinal fields and higher-order transverse fields become more substantial the smaller $kw_0$ becomes.

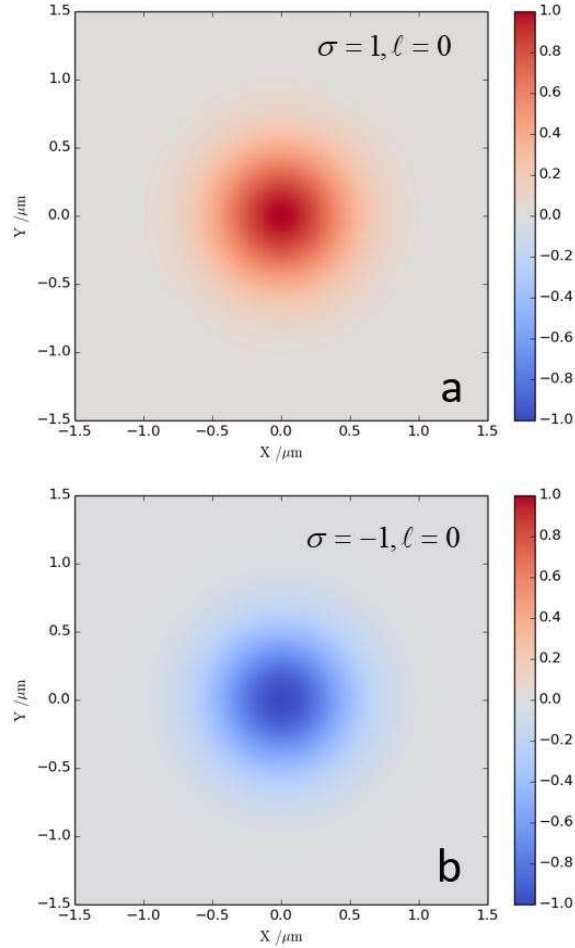

**Figure 1**: normalised optical helicity (Chirality) density distribution in the focal plane for $w_0 = \lambda$ for $\ell = 0$ and a) $\sigma = 1$ b) $\sigma = -1$.

The other two distinct combinations are for parallel $\text{sgn}\,\sigma = \text{sgn}\,\ell$ or anti-parallel $\text{sgn}\,\sigma = -\text{sgn}\,\ell$ combinations of SAM and OAM. The most interesting examples of the difference in behaviour of these two distinct combinations are for the $|\ell| = 1$ modes, and these are plotted in Figure 2 (all Figures throughout this work are plotted for $p = 0$). We see that when $\text{sgn}\,\sigma = -\text{sgn}\,\ell$ we produce an on-axis optical helicity density in the focal plane, this SOI is also responsible for the on-axis intensity of tightly-focused vortex beams [50,51]. The same mechanism is at play for $|\ell| = 2$ - plotted in Figure 3 – however the effect is much less obvious.

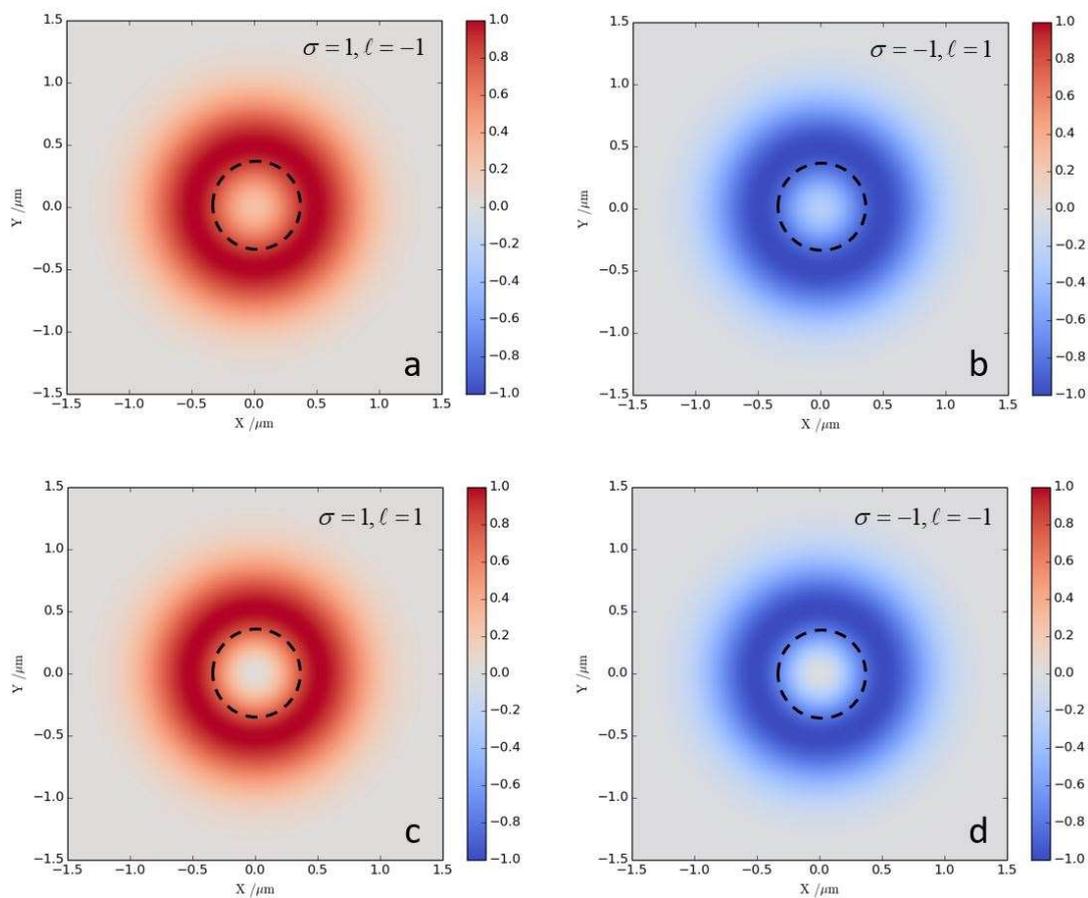

**Figure 2**: normalised optical helicity (Chirality) density distribution in the focal plane for $w_0 = \lambda$ a) $\sigma = 1, \ell = -1$ b) $\sigma = -1, \ell = 1$ c) $\sigma = 1, \ell = 1$ d) $\sigma = -1, \ell = -1$. Dashed circles aid clarity of differences. $p = 0$ (a)-(d).

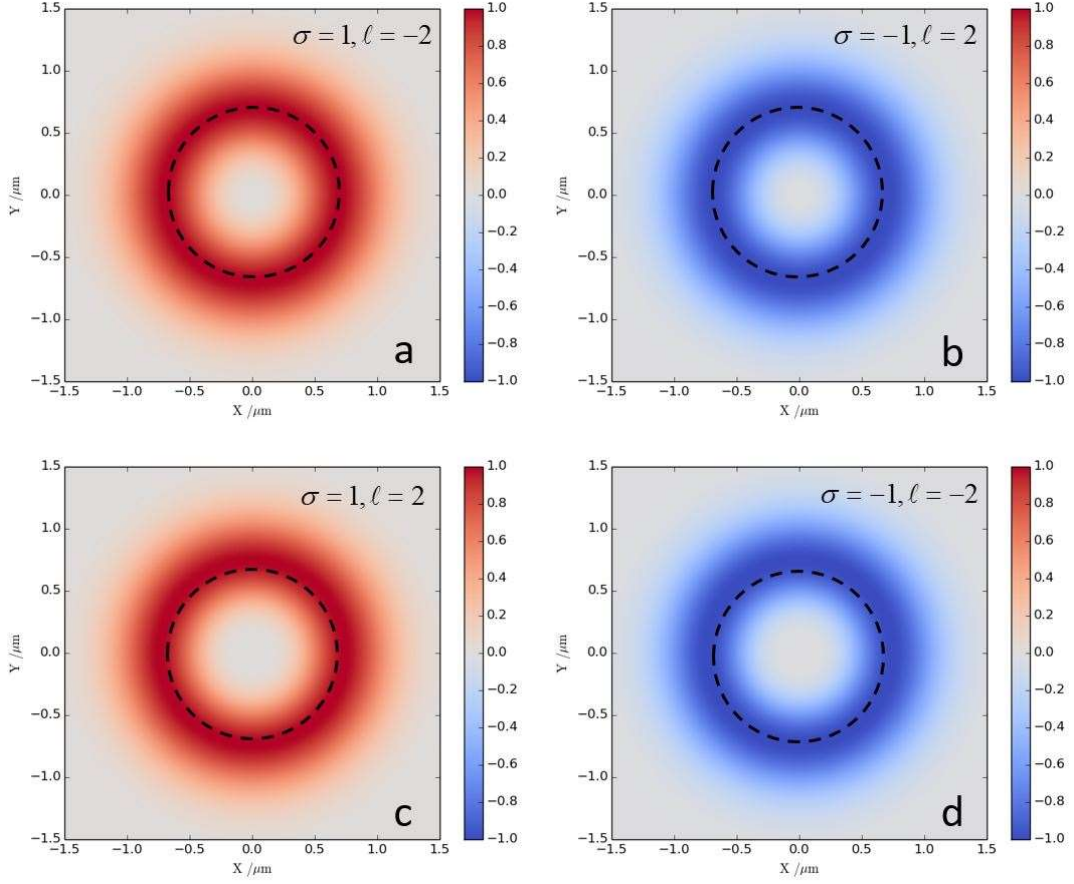

**Figure 3**: normalised optical helicity (Chirality) density distribution in the focal plane for $w_0 = \lambda$ a) $\sigma = 1, \ell = -2$ b) $\sigma = -1, \ell = 2$ c) $\sigma = 1, \ell = 2$ d) $\sigma = -1, \ell = -2$. Dashed circles aid clarity of differences. $p = 0$ (a)-(d).

### 4.3 *linear polarisation*

For a linearly polarised beam in the *x*-direction the field mode operators take the form;

$$e^{\perp}(\mathbf{r}) = i \sum_{k,\ell,p} \left( \frac{\hbar c k}{2\varepsilon_0 A_{\ell,p}^2 V} \right)^{1/2} \left[ \hat{\mathbf{x}} + \hat{\mathbf{z}} \frac{i}{k} \left\{ (\cos\phi) \frac{\partial}{\partial r} - \frac{i\ell}{r}(\sin\phi) \right\} \right] \\ \times a_{\ell,p}(k\hat{\mathbf{z}}) f \, e^{i\ell\phi} \, e^{ikz} - \text{H.c} \Big], \qquad (15)$$

and

$$\mathbf{b}(\mathbf{r}) = i \sum_{k,\ell,p} \left( \frac{\hbar k}{2c\varepsilon_0 A_{\ell,p}^2 V} \right)^{1/2} \left[ \hat{\mathbf{y}} + \hat{\mathbf{z}} \frac{i}{k} \left\{ (\sin\phi) \frac{\partial}{\partial r} + \frac{i\ell}{r}(\cos\phi) \right\} \right] \\ \times a_{\ell,p}(k\hat{\mathbf{z}}) f \, e^{i\ell\phi} \, e^{ikz} - \text{H.c} \Big]. \qquad (16)$$

Once again, inserting (15) and (16) gives an optical helicity density

$$h = -\sum_{\ell,p}\left(\frac{n\hbar}{A_{\ell,p}^2 V}\right)\frac{1}{k^2}\frac{\ell}{r}ff', \qquad (17)$$

which we see has the correct units of an angular momentum density. The helicity density distribution (17) is shown for $\ell = \pm 1, \pm 2$ in Figures 4 and 5, respectively. We observe a non-zero optical helicity (and optical chirality) for a linear-polarised electromagnetic field, for the $\ell = \pm 1$ case the maximum value is on-axis along the so-called optical vortex core. This feature of optical vortices was first pointed out by Rosales-Guzmán *et al.* for Bessel beams [52] and subsequently experimentally verified for Laguerre-Gaussian beams by P. Woźniak *et al.* [53]. We return to this result in the Discussion Section 7. A more recent study also looked at the optical chirality and helicity of linearly-polarised Laguerre-Gaussian modes [54] but produced results that do not match our results or the previous studies – we return to this issue in Section 5.4.

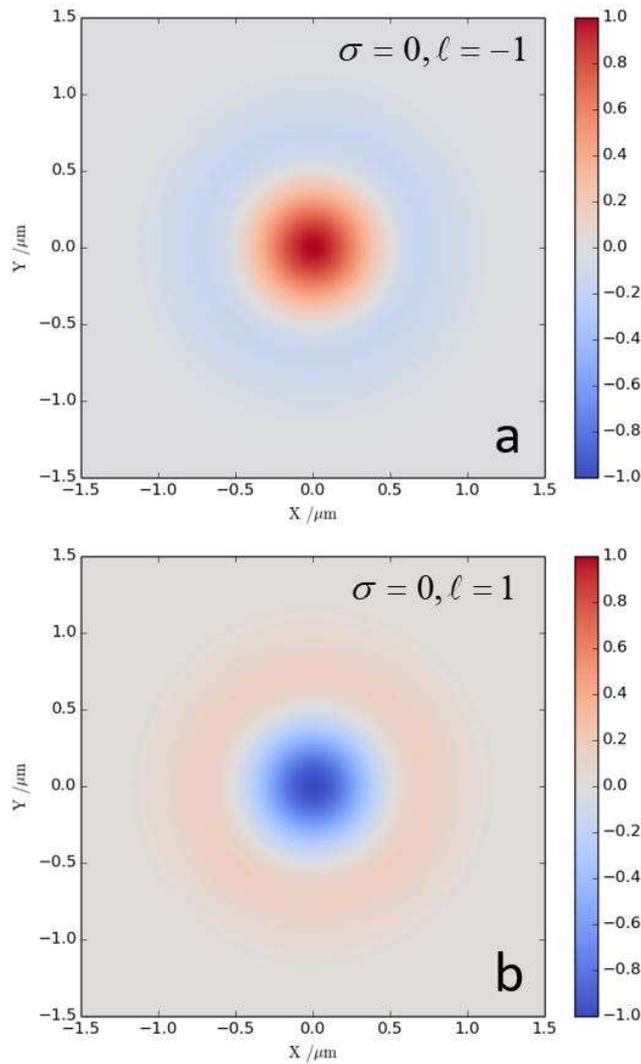

11**Figure 4**: normalised optical helicity (Chirality) density distribution in the focal plane for $w_0 = \lambda$ a) $\sigma = 0, \ell = -1$ b) $\sigma = 0, \ell = 1$. $p = 0$ (a)-(b).

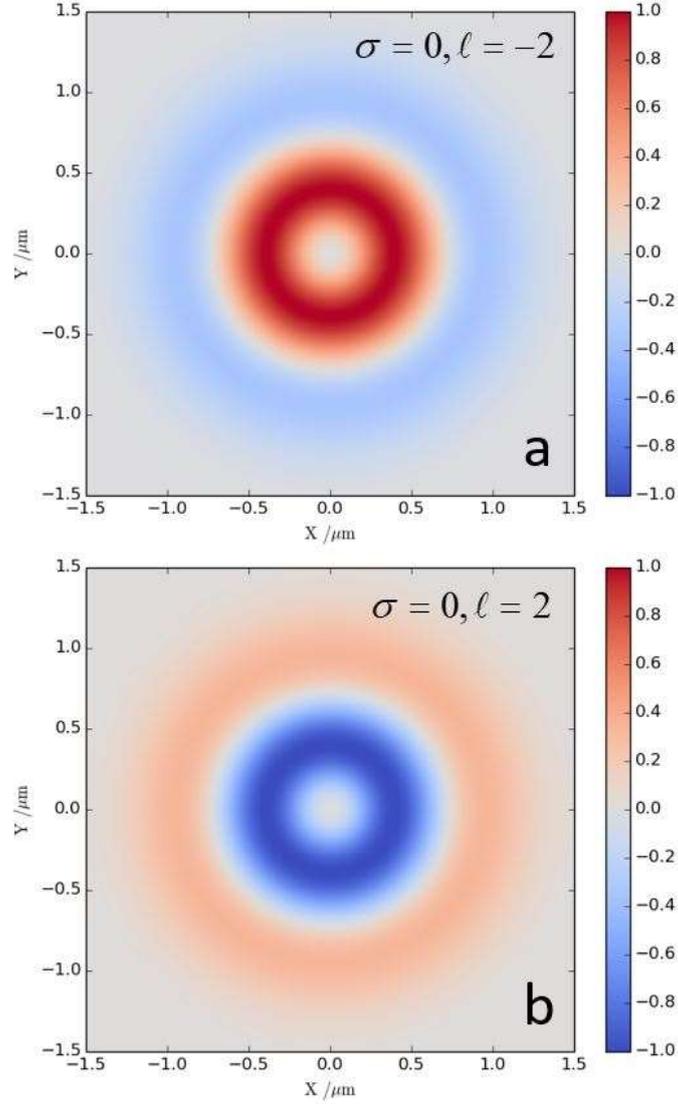

**Figure 5**: normalised optical helicity (Chirality) density distribution in the focal plane for $w_0 = \lambda$ a) $\sigma = 0, \ell = -2$ b) $\sigma = 0, \ell = 2$. $p = 0$ (a)-(b).

Unlike the optical helicity $\mathcal{H}$ of a circularly-polarised monochromatic beam which has a non-zero value, the relevant integral of (17) is zero:

$$\mathcal{H} = \int h d\mathbf{r}^2 = 0. \tag{18}$$

This indicates that particles smaller than the beam waist may probe the optical helicity density (17), but particles which are larger couple to (18) and thus do not exhibit the propensity to engage with the optical helicity generated from OAM in the focal plane.

Sections 4.2 and 4.3 show us that the optical helicity and optical chirality are influenced by orbital angular momentum. Thus we see that the conclusions of Cole and Andrews [55,56], i.e. OAM does not influence optical helicity (Chirality), are only correct when restricted to the zeroth-order transverse components of the field and is in general not true for optical vortices. Of course, the integrated value $\mathcal{H}$ is purely a measure of the number of circularly polarised photons, but the optical helicity (and Chirality) *densities* are significantly influenced by optical OAM of vortex beams. Qualitatively we see the influence of optical OAM on these quantities most dramatically for linearly polarised inputs.

## 5. Contributions involving the second-order transverse fields

**5.1**

When looking at (11) the terms $T_0^e \cdot T_2^b + T_0^b \cdot T_2^e$ are theoretically of the same magnitude as the $L_1^e \cdot L_1^b$ term of the previous Section. As such, their contributions should also be calculated and taken account of. Similarly, we carry out this analysis for a circular- and linearly polarised input LG mode.

### 5.2 *circular polarisation*

The electromagnetic field mode expansions operators now include the $T_2^{e/b}$ components necessary to computer the second-order transverse components to the optical helicity and Chirality. For circularly polarised fields they take the form

$$\begin{aligned}
e^\perp(r) = i \sum_{k,\sigma,\ell,p} \left( \frac{\hbar c k}{2\varepsilon_0 A_{\ell,p}^2 V} \right)^{1/2} \frac{1}{\sqrt{2}} &\left[ \left( (\hat{x} + i\sigma\hat{y}) + \hat{z}\frac{i}{k}e^{i\sigma\phi}\left(\frac{\partial}{\partial r} - \frac{\ell\sigma}{r}\right) \right) \right. \\
&+ \frac{i}{k^2}e^{i\sigma\phi}\left( \hat{x}\left\{ \sigma(\sin\phi)\frac{\partial}{\partial r}\left(\frac{\partial}{\partial r}\right) + \frac{i(\ell\sigma+1)}{r}(\cos\phi)\left(\frac{\partial}{\partial r}\right) - (\sin\phi)\left(\frac{\partial}{\partial r}\frac{\ell}{r}\right) - \frac{i(\ell^2+\ell\sigma)}{r^2}(\cos\phi) \right\} \right. \\
&\left. - \hat{y}\left\{ \sigma(\cos\phi)\frac{\partial}{\partial r}\left(\frac{\partial}{\partial r}\right) - \frac{i(\ell\sigma+1)}{r}(\sin\phi)\left(\frac{\partial}{\partial r}\right) - (\cos\phi)\left(\frac{\partial}{\partial r}\frac{\ell}{r}\right) + \frac{i(\ell^2+\ell\sigma)}{r^2}(\sin\phi) \right\} \right) \\
&\left. \times a_{\ell,p}^{(\sigma)}(k\hat{z}) f e^{i\ell\phi} e^{ikz} - H.c. \right],
\end{aligned} \quad (19)$$

and

$$\boldsymbol{b}(\boldsymbol{r}) = i \sum_{k,\sigma,\ell,p} \left( \frac{\hbar k}{2c\varepsilon_0 A_{\ell,p}^2 V} \right)^{1/2} \frac{1}{\sqrt{2}} \left[ (\hat{\boldsymbol{y}} - i\sigma\hat{\boldsymbol{x}}) + \hat{\boldsymbol{z}} \frac{1}{k} e^{i\sigma\phi} \left( \sigma \frac{\partial}{\partial r} - \ell \frac{1}{r} \right) \right.$$

$$-\frac{1}{k^2} e^{i\sigma\phi} \left\{ \hat{\boldsymbol{y}} \left\{ (\cos\phi) \frac{\partial}{\partial r} \left( \frac{\partial}{\partial r} \right) - \frac{i(\ell+\sigma)}{r} (\sin\phi) \frac{\partial}{\partial r} - (\cos\phi) \left( \frac{\partial}{\partial r} \frac{\ell\sigma}{r} \right) + \frac{i(\ell^2\sigma+\ell)}{r^2} (\sin\phi) \right\} \right.$$

$$\left. -\hat{\boldsymbol{x}} \left\{ (\sin\phi) \frac{\partial}{\partial r} \left( \frac{\partial}{\partial r} \right) + \frac{i(\ell+\sigma)}{r} (\cos\phi) \frac{\partial}{\partial r} - (\sin\phi) \left( \frac{\partial}{\partial r} \frac{\ell\sigma}{r} \right) - \frac{i(\ell^2\sigma+\ell)}{r^2} (\cos\phi) \right\} \right)$$

$$\times a_{\ell,p}^{(\sigma)} (k\hat{\boldsymbol{z}}) f e^{i\ell\phi} e^{ikz} - H.c. \Big]. \tag{20}$$

Inserting (19) and (20) into (6) and after some tedious algebra we produce

$$h = \sum_{\sigma,\ell,p} \left( \frac{n\hbar}{A_{\ell,p}^2 V} \right) \left[ \underbrace{\sigma f^2}_{T_0} + \frac{1}{k^2} \underbrace{\left( \frac{\sigma}{2} f'^2 - \frac{\ell}{r} f f' + \frac{\ell^2 \sigma}{2r^2} f^2 \right)}_{L_1} \right.$$

$$\left. \underbrace{-\frac{(\ell+\sigma)}{r} f f' + \frac{(\ell^2\sigma+\ell)}{r^2} f^2 - \sigma f f'' + f \frac{\partial}{\partial r} \frac{\ell}{r} f}_{T_2} \right], \tag{21}$$

where we have specifically labelled what parts can be attributed to what order fields. We see (21) has the correct units of an angular momentum density. The difference between (21) and (14) is that the former includes the $T_2^{e/b}$ components and in Figure 7 the radial distributions of the optical helicity for both has been plotted against one another, highlighting that including $T_2^{e/b}$ does not affect the on-axis optical helicity of $\ell \neq 0$ beams, but adds quantitative corrections to the transverse spatial distributions.

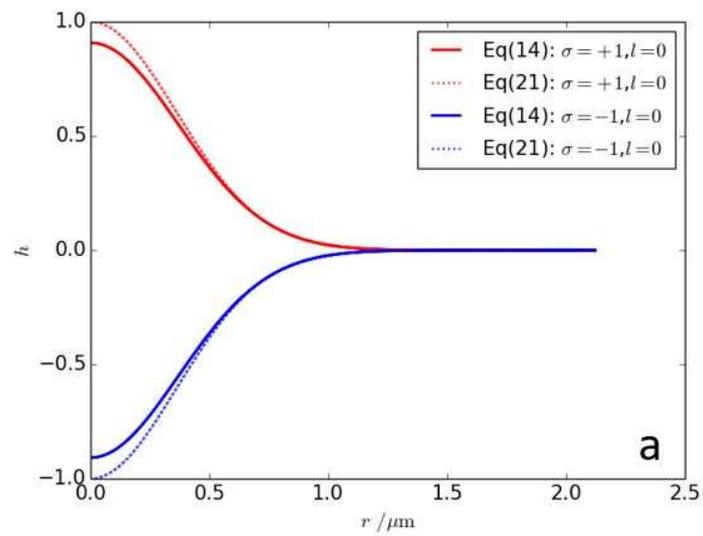

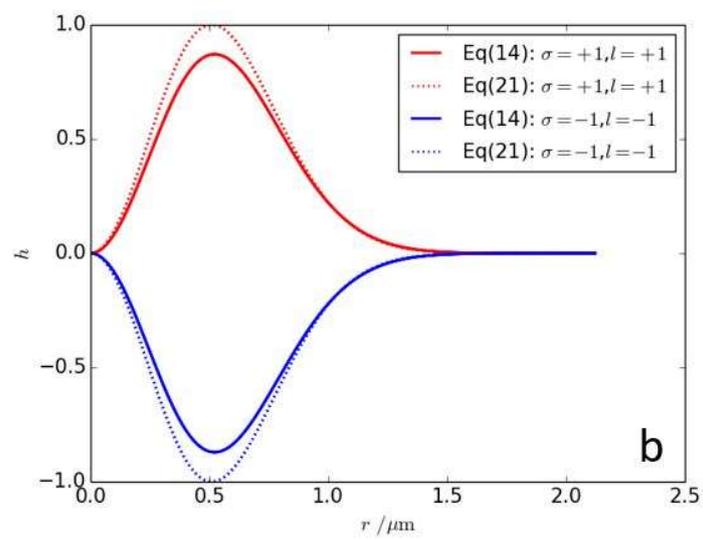

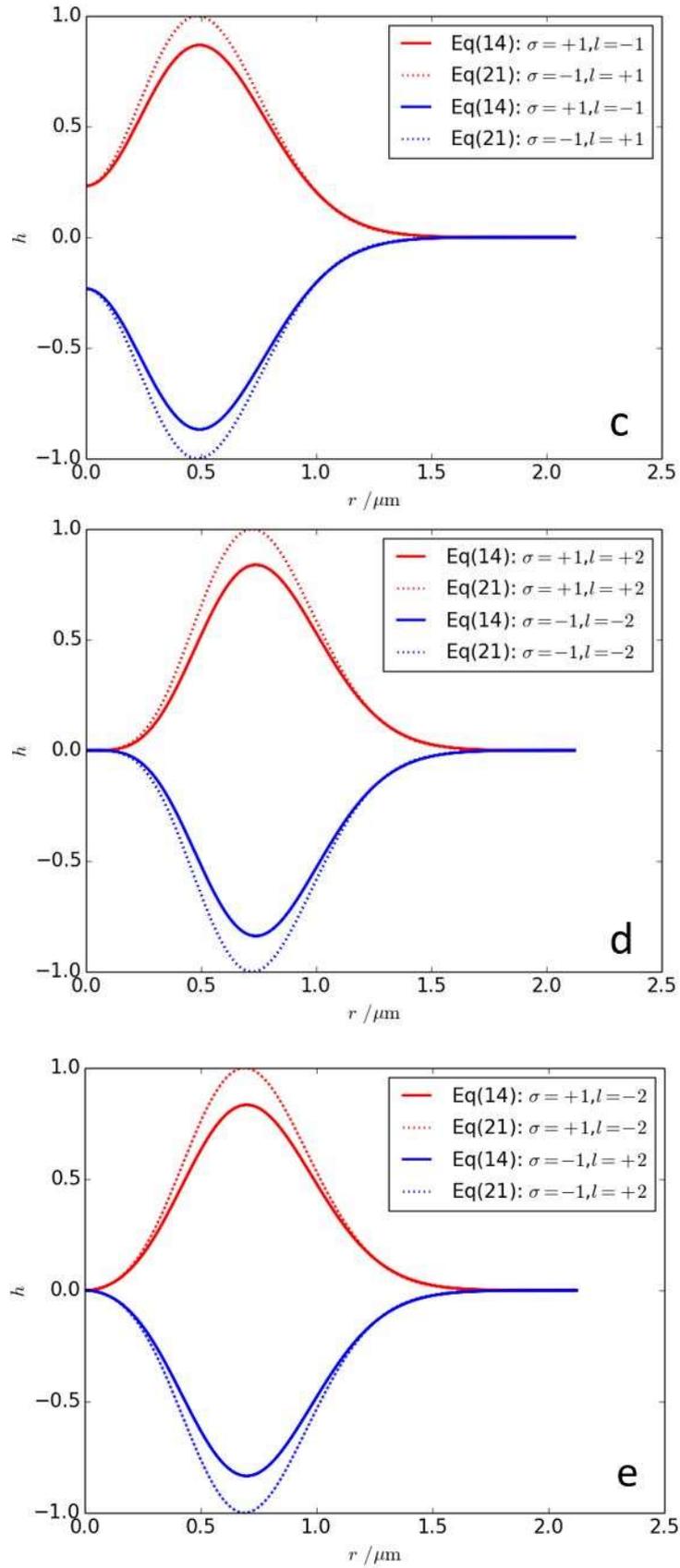

**Figure 6**: normalised optical helicity (Chirality) density distribution in the focal plane for $w_0 = \lambda$ a) $\sigma = 0, \ell = -2$ b) $\sigma = 0, \ell = 2$. $p = 0$. note that the solid lines are the line plots of the transverse spatial distributions in Figures 1-3.

### 5.3 *linear polarisation*

The electromagnetic field mode expansions operators including the $T_2^{e/b}$ components for linearly polarised in the *x*-direction inputs are

$$e^{\perp} = i\sum_{k,\ell,p}\left(\frac{\hbar ck}{2\varepsilon_0 A_{\ell,p}^2 V}\right)^{1/2}\left[\left(\hat{x}+\frac{i}{k}\hat{z}\left\{\cos\phi\frac{\partial}{\partial r}-\frac{i\ell}{r}\sin\phi\right\}\right.\right.$$
$$+\frac{1}{k^2}\left(\hat{y}\left\{\cos\phi\sin\phi\frac{\partial}{\partial r}\left(\frac{\partial}{\partial r}\right)-\frac{i\ell}{r}\sin^2\phi\frac{\partial}{\partial r}-\frac{1}{r}\cos\phi\sin\phi\frac{\partial}{\partial r}+\cos^2\phi\left(\frac{\partial}{\partial r}\frac{i\ell}{r}\right)+\frac{\ell^2}{r^2}\cos\phi\sin\phi+\frac{i\ell}{r^2}\sin^2\phi\right\}\right.$$
$$+\hat{x}\left\{-\sin^2\phi\frac{\partial}{\partial r}\left(\frac{\partial}{\partial r}\right)-\frac{i\ell}{r}\sin\phi\cos\phi\frac{\partial}{\partial r}-\frac{1}{r}\cos^2\phi\frac{\partial}{\partial r}-\cos\phi\sin\phi\left(\frac{\partial}{\partial r}\frac{i\ell}{r}\right)+\frac{\ell^2}{r^2}\cos^2\phi+\frac{i\ell}{r^2}\sin\phi\cos\phi\right\}\right)\right]$$
$$\times a_{\ell,p}(k\hat{z})fe^{i\ell\phi}e^{ikz}-\text{H.c.}\bigg], \tag{22}$$

and

$$b = i\sum_{k,\ell,p}\left(\frac{\hbar k}{2c\varepsilon_0 A_{\ell,p}^2 V}\right)^{1/2}\left[\left(\hat{y}+\frac{i}{k}\hat{z}\left\{\sin\phi\frac{\partial}{\partial r}+\frac{i\ell}{r}\cos\phi\right\}\right.\right.$$
$$+\frac{1}{k^2}\left(\hat{x}\left\{\sin\phi\cos\phi\frac{\partial}{\partial r}\left(\frac{\partial}{\partial r}\right)+\frac{i\ell}{r}\cos^2\phi\frac{\partial}{\partial r}-\frac{1}{r}\sin\phi\cos\phi\frac{\partial}{\partial r}-\sin^2\phi\left(\frac{\partial}{\partial r}\frac{i\ell}{r}\right)+\frac{\ell^2}{r^2}\sin\phi\cos\phi-\frac{i\ell}{r^2}\cos^2\phi\right\}\right.$$
$$+\hat{y}\left\{-\cos^2\phi\frac{\partial}{\partial r}\left(\frac{\partial}{\partial r}\right)+\frac{i\ell}{r}\cos\phi\sin\phi\frac{\partial}{\partial r}-\frac{1}{r}\sin^2\phi\frac{\partial}{\partial r}+\cos\phi\sin\phi\left(\frac{\partial}{\partial r}\frac{i\ell}{r}\right)+\frac{\ell^2}{r^2}\sin^2\phi-\frac{i\ell}{r^2}\cos\phi\sin\phi\right\}\right)\right]$$
$$\times a_{\ell,p}(k\hat{z})fe^{i\ell\phi}e^{ikz}-\text{H.c.}\bigg]. \tag{23}$$

The optical helicity density is

$$h = \sum_{\ell,p}\left(\frac{n\hbar}{A_{\ell,p}^2 V}\right)\frac{1}{k^2}\left[\frac{\ell}{r^2}f^2+f\frac{\partial}{\partial r}\frac{\ell}{r}f-\frac{2\ell}{r}ff'\right]. \tag{24}$$

We see that including linearly polarised second-order transverse fields to the order of $(kw_0)^{-2}$ does not affect the optical helicity (Chirality) of a laser beam because

$$f\frac{\partial}{\partial r}\frac{\ell}{r}f = -\frac{\ell}{r^2}f^2+\frac{\ell}{r}ff', \tag{25}$$

which put back into (24) gives the same result as eq. (17) which was derived by including only the zeroth-order transverse and first-order longitudinal components of the field: the optical helicity (and optical chirality) that stems from the OAM is manifest purely through longitudinal components of the field. Furthermore, whilst the helicity associated with circular-polarisation of transverse fields correlates to a spin density in the direction of propagation, we see that the helicity associated with longitudinal fields is related to the *transverse* spin density [57,58] (See Section 6).

### *5.4 Comment on Köksal et al. study*

A recent study by Köksal *et al*. [54] examined the optical chirality and helicity for monochromatic linearly polarised (in the ***x***-direction) Laguerre-Gaussian beams in free space. However, although an identical system to ours, their results conflict with those given here, as well as those in previous studies [49,52,53]. In their work they addressed why their results differ from those given in [53] specifically, suggesting that because the light is focused by a lens in the experiment this alters the optical chirality and helicity and their theory does not account for influence of the lens apparatus. However, in this study we have not considered the act of any experimental apparatus explicitly and our results fully agree with previous theory and qualitatively with experiment. In fact, both Köksal *et al*. and the work here relies on the magnitude of the paraxial factor $kw_0$, i.e. the ratio of wavelength to beam waist, which essentially accounts for focusing. The reason why the Köksal *et al*. study does not produce the same results as other studies and here is that they do not include all the necessary terms in their calculations. Specifically, in their study they include the $\chi \propto \boldsymbol{T}_0^e \cdot \boldsymbol{T}_0^b + (kw_0)^{-2}\left(\boldsymbol{T}_0^e \cdot \boldsymbol{T}_2^b + \boldsymbol{L}_1^e \cdot \boldsymbol{L}_1^b\right)$ terms of (11) but have neglected the term $\boldsymbol{T}_0^b \cdot \boldsymbol{T}_2^e$. If we go back to our calculations and neglect the same term Köksal *et al*. do then we produce the following optical chirality density

$$\chi' = -\sum_{\ell,p}\left(\frac{n\hbar ck^2}{A_{\ell,p}^2 V}\right)\frac{1}{k^2}\left[\frac{\ell}{r}\left(\cos^2\phi\right)ff' - \left(\sin^2\phi\right)f\frac{\partial}{\partial r}\frac{\ell}{r}f - \frac{\ell}{r^2}\left(\cos^2\phi\right)f^2 + \frac{\ell}{r}ff'\right], \qquad (26)$$

where the first three terms in square brackets stem from the transverse fields and the final term is that from the longitudinal contribution. Plotting (26) for $\ell = \pm 1, p = 0$ produces qualitatively the same optical chirality density distributions in the focal plane as given in Figure 2 in the paper by Köksal *et al*. We note that our minima are not the same absolute magnitude as our maxima as is the case in Köksal *et al*.

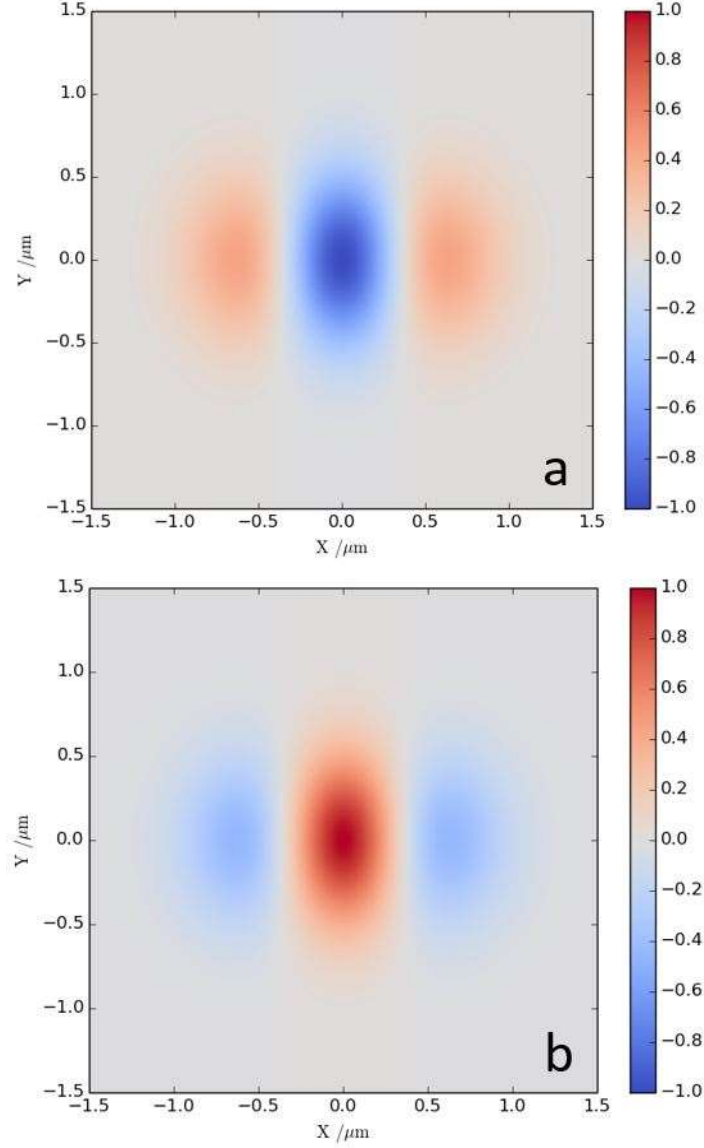

**Figure 7**: Plot of Eq. (26) which produces the same optical chirality density (qualitatively – see main text) in the focal plane as the Köksal *et al.* study (Figure 2 of Ref [54]). a) $\ell = 1 \ p = 0$ b) $\ell = -1 \ p = 0$.

The term $\boldsymbol{T}_0^b \cdot \boldsymbol{T}_2^e$ neglected by Köksal *et al*. is given explicitly as

$$\chi\left(\mathrm{T}_0^b \mathrm{T}_2^e\right) = -\sum_{k,\ell,p}\left(\frac{n\hbar ck^2}{A_{\ell,p}^2 V}\right)\ \frac{1}{k^2}\left[\frac{\ell}{r}\left(\sin^2\phi\right)ff' - \left(\cos^2\phi\right)f\frac{\partial}{\partial r}\frac{\ell}{r}f - \frac{\ell}{r^2}\left(\sin^2\phi\right)f^2\right], \tag{27}$$

which when added to (26) gives the circularly-symmetric

$$\chi = \sum_{\ell,p} \left( \frac{n\hbar c k^2}{A_{\ell,p}^2 V} \right) \frac{1}{k^2} \left[ \frac{\ell}{r^2} f^2 + f \frac{\partial}{\partial r} \frac{\ell}{r} f - \frac{2\ell}{r} f f' \right], \qquad (28)$$

and is essentially what is given in (24) besides the factor of $ck^2$, i.e. $h = \chi/ck^2$, with the corresponding circular-symmetric optical chirality distributions of Figures 4 and 5. As such, by including the term $\boldsymbol{T}_0^b \cdot \boldsymbol{T}_2^e$ we produce the expected result.

The origin of the importance of including all the necessary terms is that the electric and magnetic fields of radiation fields do not necessarily contribute equally to conserved quantities in free space. What we observe here is that $\boldsymbol{T}_0^b \cdot \boldsymbol{T}_2^e \neq \boldsymbol{T}_2^b \cdot \boldsymbol{T}_0^e$ for a linear polarised field, however this doesn't suggest that we require a specific dual-symmetric equation to calculate the helicity and Chirality: as noted in Section 2 the equation for optical helicity and Chirality is identical in both symmetric and asymmetric formulations [37] (The helicity and optical chirality are unique quantities in this respect for the free field). Rather all terms must be included up to a given order of the paraxial parameter. Optical helicity and optical chirality are in fact particularly unique as unlike other quantities (canonical momentum density, spin momentum density, energy density, etc.) which engage with the electric biased nature of materials in actual experiments (generally), optical helicity and optical chirality specifically couple to chiral matter via electric and magnetic dipoles, and so this is why the experiment produces the circularly-symmetric helicity rather than the electric biased result (26).

### 6. Spin angular momentum density

As mentioned in Section 2, there is a continuity equation (4) which relates helicity to spin. The spin angular momentum density $\boldsymbol{s}$ is given explicitly in coulomb gauge QED as

$$\boldsymbol{s} = \frac{\varepsilon_0}{2} \left( -\boldsymbol{e}^\perp \times \int \boldsymbol{e}^\perp dt - \frac{1}{c} \boldsymbol{b} \times \int \boldsymbol{b} dt \right). \qquad (29)$$

As shown for the optical helicity and Chirality, the spin density can similarly be calculated to an $n$-th order in the paraxial parameter:

$$\begin{aligned}
\boldsymbol{s} &\propto \boldsymbol{T}_0^{e/b} \times \boldsymbol{T}_0^{e/b}, \\
\boldsymbol{s}' &\propto \boldsymbol{T}_0^{e/b} \times \boldsymbol{T}_0^{e/b} + \boldsymbol{T}_0^{e/b} \times \boldsymbol{L}_1^{e/b} \left( \frac{1}{kw_0} \right), \\
\boldsymbol{s}'' &\propto \boldsymbol{T}_0^{e/b} \times \boldsymbol{T}_0^{e/b} + \boldsymbol{T}_0^{e/b} \times \boldsymbol{L}_1^{e/b} \left( \frac{1}{kw_0} \right) + \boldsymbol{T}_0^{e/b} \times \boldsymbol{T}_2^{e/b} \left( \frac{1}{(kw_0)^2} \right).
\end{aligned} \qquad (30)$$

Unlike the optical helicity and Chirality (as well as energy density, for example) which progress in $2n$ degrees of the paraxial parameter (see (11)) the spin angular momentum density involves $n$ contributions. The zeroth-order term $\boldsymbol{T}_0^{e/b} \times \boldsymbol{T}_0^{e/b}$ is zero for linearly polarised beams, and its integral value takes on values of $\pm\hbar$ per photon in the direction of propagation for circular polarisation. The first-order contribution is responsible for transverse spin density [57–59], calculated with our QED mode expansions as:

$$\boldsymbol{T}_0^{e/b} \times \boldsymbol{L}_1^{e/b}\left(\frac{1}{kw_0}\right) = \sum_{k,\ell,p}\left(\frac{n\hbar}{A_{\ell,p}^2 V}\right)\left[\underbrace{\hat{\boldsymbol{x}}\frac{1}{k}(\sin\phi)\frac{\partial}{\partial r}}_{s_x'^b} - \underbrace{\hat{\boldsymbol{y}}\frac{1}{k}(\cos\phi)\frac{\partial}{\partial r}}_{s_y'^e}\right]ff$$

$$= -\hat{\boldsymbol{\phi}}\sum_{k,\ell,p}\left(\frac{n\hbar}{A_{\ell,p}^2 V}\right)\frac{1}{k}ff', \quad (31)$$

where $s_x'^b$ is the magnetic field contribution and $s_y'^e$ is the electric field contribution [60]. The individual $s_x'^b$ and $s_y'^e$, as well as the total $s_\phi^{e+b}$, parts of (31) are plotted in Figure 8 for $\ell=1, p=0, w_0=\lambda$.

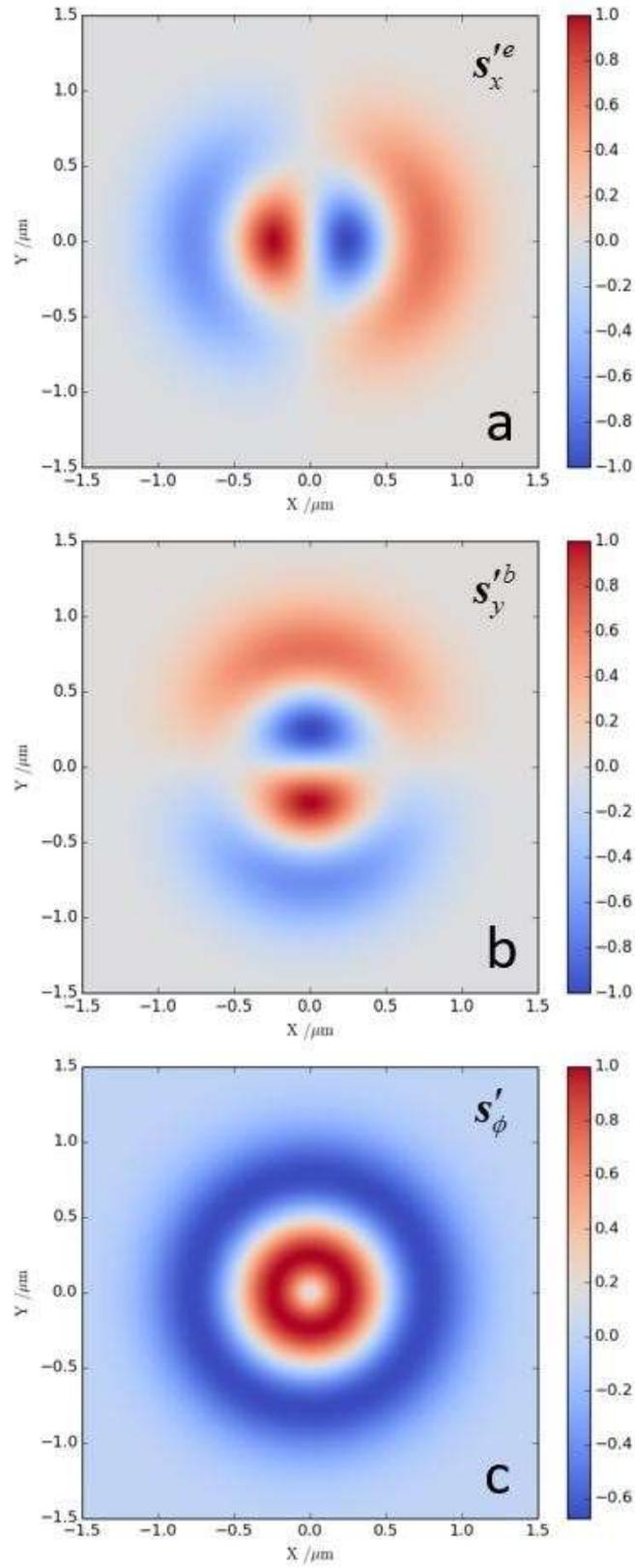

**Figure 8**: Components of transverse SAM density (a) $s_y'^e$ term in Eq. (31) ; (b) $s_x'^b$ term in Eq. (31); and (c) the total $s_\phi'^{e+b}$ Eq. (31). $|\ell|=1, p=0, w_0=\lambda$ for (a)-(c).

The second order contribution using QED mode expansions is:

$$\boldsymbol{T}_0^{e/b} \times \boldsymbol{T}_2^{e/b} \left( \frac{1}{(kw_0)^2} \right) = \sum_{k,\ell,p} \left( \frac{n\hbar}{A_{\ell,p}^2 V} \right) \frac{2}{k^2} \hat{z} \left[ \underbrace{\left\{ \cos^2\phi \left( \frac{\partial}{\partial r} \frac{\ell}{r} \right) + \frac{\ell}{r^2} \sin^2\phi - \frac{\ell}{r} \frac{\partial}{\partial r} \sin^2\phi \right\}}_{s_z^{\prime\prime e}} \right.$$

$$\left. + \underbrace{\left\{ \sin^2\phi \left( \frac{\partial}{\partial r} \frac{\ell}{r} \right) + \frac{\ell}{r^2} \cos^2\phi - \frac{\ell}{r} \frac{\partial}{\partial r} \cos^2\phi \right\}}_{s_z^{\prime\prime b}} \right] ff$$

$$= \sum_{k,\ell,p} \left( \frac{n\hbar}{A_{\ell,p}^2 V} \right) \frac{2}{k^2} \hat{z} \left[ \left( \frac{\partial}{\partial r} \frac{\ell}{r} \right) + \frac{\ell}{r^2} - \frac{\ell}{r} \frac{\partial}{\partial r} \right] ff. \quad (32)$$

Remarkably what this shows is that there exists a SAM density in the direction of propagation even for linearly polarised input optical vortices [61–63]. Note that each term in (32) is dependent on $\ell$ and so this phenomena is unique to optical vortices: a linearly polarised Gaussian beam $\ell = 0$ does not possess this longitudinal SAM density. The total dual symmetric contribution (32) is actually zero due to (25), however experimentally of course light-matter interactions are generally dominated by electric dipole coupling and so the electric field contribution to (32) should be observable provided $kw_0$ is small enough. The individual electric $s_z^{\prime\prime e}$ and magnetic $s_z^{\prime\prime b}$ contributions of Eq. (32) are plotted in Fig. 9.

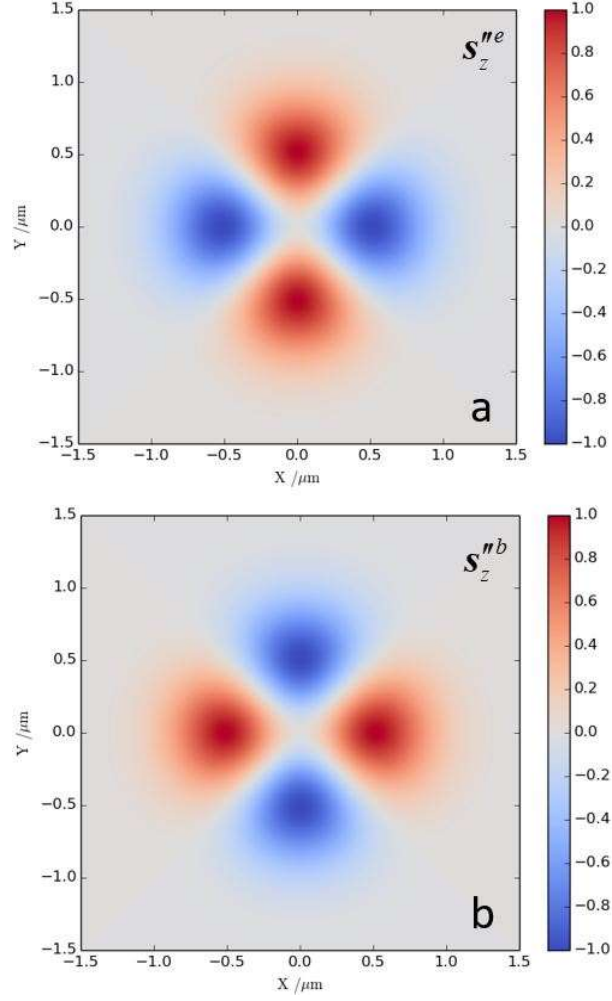

**Figure 9**: Components of longitudinal SAM density of (a) $s_z''^{e}$ term in Eq. (32); (b) $s_z''^{b}$ term in Eq.(32) $|\ell|=1, p=0, w_0 = \lambda$ for (a)-(b).

Although the optical helicity was calculated using the formula (6), another way is to calculate the projection of the spin onto the linear momentum density $s \cdot p_\text{O}/p_\text{O}(z)$, where $p_\text{O}$ is the (orbital) canonical momentum density, calculated in paraxial form as

$$p_\text{O} = \frac{\varepsilon_0}{2}\left[e^\perp \cdot \nabla a^\perp + b \cdot \nabla c^\perp\right]$$
$$= \sum_{k,\ell,p}\left(\frac{n\hbar}{A_{\ell,p}^2 V}\right)\left(\frac{\ell}{r}\hat{\phi} + k\hat{z}\right)ff^*. \qquad (33)$$

It is easily seen that projecting (31) on to (33) gives the correct helicity density (17) which was calculated using the integrand of (3). The helicities associated with the second order spin density (32) are

$$h''\left(s_z^{n'e}\right) = \sum_{k,\ell,p}\left(\frac{n\hbar}{A_{\ell,p}^2 V}\right)\frac{2}{k^2}\left\{\cos^2\phi\left(\frac{\partial}{\partial r}\frac{\ell}{r}\right) + \frac{\ell}{r^2}\sin^2\phi - \frac{\ell}{r}\frac{\partial}{\partial r}\sin^2\phi\right\}ff$$

$$= \sum_{k,\ell,p}\left(\frac{n\hbar}{A_{\ell,p}^2 V}\right)\frac{2}{k^2}\left\{\frac{\ell}{r}\frac{\partial}{\partial r} - \frac{\ell}{r^2}\right\}ff\cos 2\phi, \quad (34)$$

$$h''\left(s_z^{n'b}\right) = \sum_{k,\ell,p}\left(\frac{n\hbar}{A_{\ell,p}^2 V}\right)\frac{2}{k^2}\hat{z}\left\{\sin^2\phi\left(\frac{\partial}{\partial r}\frac{\ell}{r}\right) + \frac{\ell}{r^2}\cos^2\phi - \frac{\ell}{r}\frac{\partial}{\partial r}\cos^2\phi\right\}$$

$$= \sum_{k,\ell,p}\left(\frac{n\hbar}{A_{\ell,p}^2 V}\right)\frac{2}{k^2}\left\{\frac{\ell}{r^2} - \frac{\ell}{r}\frac{\partial}{\partial r}\right\}ff\cos 2\phi. \quad (35)$$

The sum of these two contributions (which is obviously zero) to the optical helicity density correspond to those terms calculated in (24) which are zero. The relevance of these two contributions to the helicity (34) and (35) individually are not important as we have stated the material response to helicity is not electric biased and so the individual non-zero (34) and (35) cannot be observed experimentally.

### 7. Discussion and Conclusion

We now discuss in more detail the specific result Eq. (17). What this tells us is that linearly-polarised optical vortices with small beam waists on the order of the input wavelength $w_0 \approx \lambda$ acquire a significant optical helicity and Chirality density in the focal plane, even though they possess no spin, helicity, or Chirality before focusing (i.e. when $w_0 \gg \lambda$). Clearly circular polarisation and helicity (Chirality) are not synonymous. This phenomena is a spin-orbit-interaction of light (SOI) [50], but in contrast to the very well-known spin-to-orbital angular momentum conversion, this mechanism is an orbital-to-spin angular momentum density conversion. Whilst SAM-OAM conversions are very well-known and studied, the OAM-SAM conversion process has only recently seen a large degree of research activity, and these studies are have been interested in the corresponding spin-angular momentum density (Section 6) generation from OAM which leads to mechanical effects on matter [61,62,64–68] (rather than the corresponding optical helicity and Chirality density generation we have been interested in, which relate to spectroscopic light-matter interactions). The OAM-SAM conversion is more obvious to recognise by recalling the continuity equation (4) : a degree of helicity density is produced (Eq.(17)) which is accompanied by a generation spin angular momentum density $s$. We obtain this same result by calculating $s$ to begin with and then calculating the corresponding helicity density generation (Section 6). A very important point to note however is that the quantities $\chi$ and $h$ are local and intrinsic, being linked to light's polarisation. The presence of $\ell$ in (17) and many other equations throughout this work suggest the ability of optical vortices to engage in chiroptical interactions with matter via the handedness/chirality of an optical vortex. More strictly, the handedness of the optical vortex determines the sign of the quantities $\chi$ and $h$, and it is these quantities which play the role in light-matter interactions. As such, chiral matter may *indirectly* show a discriminatory response to the pseudoscalar $\ell$ [19,49,53].

This still leaves the question to whether the actual spatial (i.e. geometrical) chirality of optical vortices influences spectroscopic light-matter interactions directly through the sign of $\ell$. This has been a question researchers have been asking for quite some time now [19,69], and very recently Nechayev *et al*. [15] introduced the term 'Kelvin's Chirality' to describe this geometrical chirality that may be possessed by structured light beams, a property seemingly quite distinct from the local properties (with the introduction of Kelvin's chirality of optical beams, it may be necessary in future to refer to the distinctly different optical chirality studied in this article as 'Lipkin's chirality'). A QED study [70] highlighted how electric quadrupole couplings are essential for chiral particles to engage with the Kelvin's chirality of optical vortices (via electric-dipole electric-quadrupole interferences), the reason being they interact with the transverse gradient of the field (even in the paraxial approximation, i.e. zeroth-order transverse field components only), of which the helical part is what gives vortices their OAM of $\ell\hbar$. This initial study was specifically concerned with dichroic-like absorption of optical vortices, and further studies have highlighted this dependence on $\ell$ in both linear [71] and nonlinear [72] scattering optical activities via electric quadrupole couplings. The fact that OAM of optical vortices is transferred to electronic degrees of freedom via electric quadrupole couplings (to leading order) in bound electrons agrees with these results [73]. However, without focusing (or spatial confinement of the fields) such effects are essentially still relatively small compared to those dependent on the longitudinal phase gradient associated with the wavelength, due to the fact Kelvin's chirality of these beams is a global property of the beam directly proportional to the beam waist $w_0$. Indeed, the relatively large helical pitch of circularly-polarised light (related to the wavelength) is one of the reasons why chiroptical interactions in small chiral particles are generally small in the first place; the pitch of an optical vortex is $\arctan \ell/kr$. A recent experimental study strikingly highlighted the scale-dependent nature of Kelvin's chirality by matching the size of chiral microstructures to that of the optical vortex beam waist [74]. Furthermore, although one can picture the azimuthal phase of an optical vortex tracing out a helical path on propagation, there is no optical helicity associated with this property of electromagnetic beams as can easily been seen by projecting the orbital angular momentum onto the linear momentum [69,75].

Finally, we comment on the fact we neglect the terms proportional to $(kw_0)^{-4}$ in (11) as only in the most extreme cases of subwavelength focusing could these become important. It must be noted that inclusion of those terms would also then require $\boldsymbol{L}_1^{e/b} \cdot \boldsymbol{L}_3^{e/b}$ and $\boldsymbol{T}_0^{e/b} \cdot \boldsymbol{T}_4^{e/b}$ to be included also. It can be predicted that an on axis optical chirality and helicity density would be produced by these higher-order terms when $\ell = 2$ due to the on-axis intensity which is known to exist in this regime [51].

**Acknowledgements** KAF is grateful to the Leverhulme Trust for funding him through a Leverhulme Trust Early Career Fellowship (Grant Number ECF-2019-398). We thank David L. Andrews for helpful comments.

**Appendix A: Radial distribution function**

The radial distribution function $f_{|\ell|,p}(r)$ is

$$f_{|\ell|,p}(r) = \frac{C_p^{|\ell|}}{w_0}\left(\frac{\sqrt{2}r}{w_0}\right)^{|\ell|} e^{-\frac{r^2}{w_0^2}} L_p^{|\ell|}\left[\frac{2r^2}{w_0^2}\right], \tag{A1}$$

where the normalisation constant is given by $C_p^{|\ell|} = \sqrt{2p!/\left[\pi(p+|\ell|)!\right]}$ and $L_p^{|\ell|}$ is the generalised Laguerre polynomial of order $p$. The $p=0$ case is most experimentally utilised (and corresponds to all the figures plotted in this paper), and the differentiations from the main manuscript are given analytically in this case as

$$\left(\frac{\partial}{\partial r}\right) = f'_{|\ell|,0}(r) = \frac{\partial}{\partial r} f_{|\ell|,0}(r) = \left(\frac{|\ell|}{r} - \frac{2r}{w_0^2}\right) f_{|\ell|,0}(r); \tag{A2}$$

$$\left(\frac{\partial}{\partial r}\right)\left(\frac{\partial}{\partial r}\right) = f'_{|\ell|,0}(r) f'_{|\ell|,0}(r) = \left(\frac{|\ell|}{r} - \frac{2r}{w_0^2}\right) f_{|\ell|,0}(r)\left(\frac{|\ell|}{r} - \frac{2r}{w_0^2}\right) f_{|\ell|,0}(r) = \left(\frac{|\ell|^2}{r^2} - \frac{4|\ell|}{w_0^2} + \frac{4r^2}{w_0^4}\right) f^2_{|\ell|,0}(r); \tag{A3}$$

$$\frac{\partial}{\partial r}\left(\frac{\partial}{\partial r}\right) = \frac{\partial}{\partial r} f'_{|\ell|,0}(r) = \left(\frac{|\ell|^2}{r^2} - \frac{|\ell|}{r^2} - \frac{4|\ell|}{w_0^2} - \frac{2}{w_0^2} + \frac{4r^2}{w_0^4}\right) f_{|\ell|,0}(r); \tag{A4}$$

$$\left(\frac{\partial}{\partial r}\frac{\ell}{r}\right) = \frac{\partial}{\partial r}\left(\frac{\ell}{r} f_{|\ell|,0}(r)\right) = \left(\frac{\ell(|\ell|-1)}{r^2} - \frac{2\ell}{w_0^2}\right) f_{|\ell|,0}(r). \tag{A5}$$

More generally

$$\begin{aligned}\frac{\partial}{\partial r} f_{|\ell|,p}(r) = \frac{C_p^{|\ell|}}{w_0}\left(\frac{\sqrt{2}r}{w_0}\right)^{|\ell|} e^{-\frac{r^2}{w_0^2}} &\left\{ L_p^{|\ell|}\left[\frac{2r^2}{w_0^2}\right]\left(\frac{|\ell|}{r} - \frac{2r}{w_0^2}\right) \right.\\ &\left. + \left(\frac{2r^2}{w_0^2}\right)^{-1}\left(p L_p^{|\ell|}\left[\frac{2r^2}{w_0^2}\right] - (|\ell|+p) L_{p-1}^{|\ell|}\left[\frac{2r^2}{w_0^2}\right]\right)\right\}.\end{aligned} \tag{A6}$$

**Appendix B: T$_0$ Electromagnetic vector potentials**

The zeroth-order transverse electromagnetic vector potential mode expansions for LG beams are given as

$$\boldsymbol{a}^{\perp}(\boldsymbol{r}) = \sum_{k,\eta,\ell,p} \left(\frac{\hbar}{2\varepsilon_0 c k A_{\ell,p}^2 V}\right)^{1/2} \left[\boldsymbol{e}^{(\eta)}(k\hat{\boldsymbol{z}}) f_{|\ell|,p}(r) a_{|\ell|,p}^{(\eta)}(k\hat{\boldsymbol{z}}) e^{i(kz+\ell\phi)} + \text{H.c}\right], \quad (B1)$$

where $\nabla \times \boldsymbol{a}^{\perp}(\boldsymbol{r}) = -\dot{\boldsymbol{c}}^{\perp}(\boldsymbol{r})$, thus

$$\boldsymbol{c}^{\perp}(\boldsymbol{r}) = \sum_{k,\eta,\ell,p} \left(\frac{\hbar}{2\varepsilon_0 c^3 k A_{\ell,p}^2 V}\right)^{1/2} \left[\left(\hat{\boldsymbol{z}} \times \boldsymbol{e}^{(\eta)}(k\hat{\boldsymbol{z}})\right) f_{|\ell|,p}(r) a_{|\ell|,p}^{(\eta)}(k\hat{\boldsymbol{z}}) e^{i(kz+\ell\phi)} + \text{H.c}\right]. \quad (B2)$$

**Appendix C: Explicit forms of transverse and longitudinal field components (up to second order in the paraxial parameter)**

For circularly polarised light

$$\boldsymbol{T}_0^e = i \sum_{k,\sigma,\ell,p} \left(\frac{\hbar c k}{2\varepsilon_0 A_{\ell,p}^2 V}\right)^{1/2} \frac{1}{\sqrt{2}} \left[(\hat{\boldsymbol{x}} + i\sigma\hat{\boldsymbol{y}}) a_{\ell,p}^{(\sigma)}(k\hat{\boldsymbol{z}}) f e^{i\ell\phi} e^{ikz} - \text{H.c.}\right]; \quad (C1)$$

$$\boldsymbol{L}_1^e = i \sum_{k,\sigma,\ell,p} \left(\frac{\hbar c k}{2\varepsilon_0 A_{\ell,p}^2 V}\right)^{1/2} \frac{1}{\sqrt{2}} \left[\hat{\boldsymbol{z}} \frac{i}{k} e^{i\sigma\phi} \left(\frac{\partial}{\partial r} - \frac{\ell\sigma}{r}\right) a_{\ell,p}^{(\sigma)}(k\hat{\boldsymbol{z}}) f e^{i\ell\phi} e^{ikz} - \text{H.c.}\right]; \quad (C2)$$

$$\boldsymbol{T}_2^e = i \sum_{k,\sigma,\ell,p} \left(\frac{\hbar c k}{2\varepsilon_0 A_{\ell,p}^2 V}\right)^{1/2} \frac{1}{\sqrt{2}} \left[\frac{i}{k^2} e^{i\sigma\phi} \left(\hat{\boldsymbol{x}}\left\{\sigma(\sin\phi)\frac{\partial}{\partial r}\left(\frac{\partial}{\partial r}\right) + \frac{i(\ell\sigma+1)}{r}(\cos\phi)\left(\frac{\partial}{\partial r}\right)\right\}\right.\right.$$
$$\left.-(\sin\phi)\left(\frac{\partial}{\partial r}\frac{\ell}{r}\right) - \frac{i(\ell^2+\ell\sigma)}{r^2}(\cos\phi)\right\} - \hat{\boldsymbol{y}}\left\{\sigma(\cos\phi)\frac{\partial}{\partial r}\left(\frac{\partial}{\partial r}\right) - \frac{i(\ell\sigma+1)}{r}(\sin\phi)\left(\frac{\partial}{\partial r}\right)\right.$$
$$\left.\left.-(\cos\phi)\left(\frac{\partial}{\partial r}\frac{\ell}{r}\right) + \frac{i(\ell^2+\ell\sigma)}{r^2}(\sin\phi)\right\}\right) a_{\ell,p}^{(\sigma)}(k\hat{\boldsymbol{z}}) f e^{i\ell\phi} e^{ikz} - \text{H.c.}\right]; \quad (C3)$$

$$\boldsymbol{T}_0^b = i \sum_{k,\sigma,\ell,p} \left( \frac{\hbar k}{2c\varepsilon_0 A_{\ell,p}^2 V} \right)^{1/2} \frac{1}{\sqrt{2}} \left[ (\hat{\boldsymbol{y}} - i\sigma\hat{\boldsymbol{x}})\, a_{\ell,p}^{(\sigma)}(k\hat{z})\, f\, e^{i\ell\phi}\, e^{ikz} - H.c. \right]; \quad (C4)$$

$$\boldsymbol{L}_1^b = i \sum_{k,\sigma,\ell,p} \left( \frac{\hbar k}{2c\varepsilon_0 A_{\ell,p}^2 V} \right)^{1/2} \frac{1}{\sqrt{2}} \left[ \hat{\boldsymbol{z}} \frac{1}{k} e^{i\sigma\phi} \left( \sigma \frac{\partial}{\partial r} - \ell \frac{1}{r} \right) a_{\ell,p}^{(\sigma)}(k\hat{z})\, f\, e^{i\ell\phi}\, e^{ikz} - H.c. \right]; \quad (C5)$$

$$\boldsymbol{T}_2^b = i \sum_{k,\sigma,\ell,p} \left( \frac{\hbar k}{2c\varepsilon_0 A_{\ell,p}^2 V} \right)^{1/2} \frac{1}{\sqrt{2}} \left[ -\frac{1}{k^2} e^{i\sigma\phi} \left( \hat{\boldsymbol{y}} \left\{ (\cos\phi) \frac{\partial}{\partial r} \left( \frac{\partial}{\partial r} \right) - \frac{i(\ell+\sigma)}{r} (\sin\phi) \frac{\partial}{\partial r} \right. \right. \right.$$
$$\left. -(\cos\phi) \left( \frac{\partial}{\partial r} \frac{\ell\sigma}{r} \right) + \frac{i(\ell^2\sigma + \ell)}{r^2}(\sin\phi) \right\} - \hat{\boldsymbol{x}} \left\{ (\sin\phi) \frac{\partial}{\partial r} \left( \frac{\partial}{\partial r} \right) + \frac{i(\ell+\sigma)}{r}(\cos\phi) \frac{\partial}{\partial r} \right.$$
$$\left. \left. \left. -(\sin\phi) \left( \frac{\partial}{\partial r} \frac{\ell\sigma}{r} \right) - \frac{i(\ell^2\sigma + \ell)}{r^2}(\cos\phi) \right\} \right) a_{\ell,p}^{(\sigma)}(k\hat{z})\, f\, e^{i\ell\phi}\, e^{ikz} - H.c. \right]. \quad (C6)$$

For linearly polarised (in the **x** direction) light

$$\boldsymbol{T}_0^e = i \sum_{k,\ell,p} \left( \frac{\hbar c k}{2\varepsilon_0 A_{\ell,p}^2 V} \right)^{1/2} \left[ \hat{\boldsymbol{x}} a_{\ell,p}(k\hat{z})\, f\, e^{i\ell\phi}\, e^{ikz} - H.c. \right]; \quad (C7)$$

$$\boldsymbol{L}_1^e = i \sum_{k,\ell,p} \left( \frac{\hbar c k}{2\varepsilon_0 A_{\ell,p}^2 V} \right)^{1/2} \left[ \frac{i}{k} \hat{\boldsymbol{z}} \left\{ \cos\phi \frac{\partial}{\partial r} - \frac{i\ell}{r} \sin\phi \right\} a_{\ell,p}(k\hat{z})\, f\, e^{i\ell\phi}\, e^{ikz} - H.c. \right]; \quad (C8)$$

$$\boldsymbol{T}_2^e = i \sum_{k,\ell,p} \left( \frac{\hbar c k}{2\varepsilon_0 A_{\ell,p}^2 V} \right)^{1/2} \left[ \frac{1}{k^2} \left( \hat{\boldsymbol{y}} \left\{ \cos\phi\sin\phi \frac{\partial}{\partial r}\left(\frac{\partial}{\partial r}\right) - \frac{i\ell}{r} \sin^2\phi \frac{\partial}{\partial r} - \frac{1}{r} \cos\phi\sin\phi \frac{\partial}{\partial r} \right. \right. \right.$$
$$\left. + \cos^2\phi \left( \frac{\partial}{\partial r} \frac{i\ell}{r} \right) + \frac{\ell^2}{r^2} \cos\phi\sin\phi + \frac{i\ell}{r^2}\sin^2\phi \right\} + \hat{\boldsymbol{x}} \left\{ -\sin^2\phi \frac{\partial}{\partial r}\left(\frac{\partial}{\partial r}\right) - \frac{i\ell}{r}\sin\phi\cos\phi \frac{\partial}{\partial r} - \frac{1}{r}\cos^2\phi \frac{\partial}{\partial r} \right.$$
$$\left. \left. \left. -\cos\phi\sin\phi \left( \frac{\partial}{\partial r} \frac{i\ell}{r} \right) + \frac{\ell^2}{r^2}\cos^2\phi + \frac{i\ell}{r^2}\sin\phi\cos\phi \right\} \right) a_{\ell,p}(k\hat{z})\, f\, e^{i\ell\phi}\, e^{ikz} - H.c. \right]; \quad (C9)$$

$$\boldsymbol{T}_0^b = i \sum_{k,\ell,p} \left( \frac{\hbar k}{2c\varepsilon_0 A_{\ell,p}^2 V} \right)^{1/2} \left[ \hat{\boldsymbol{y}} a_{\ell,p}(k\hat{z})\, f\, e^{i\ell\phi}\, e^{ikz} - H.c. \right]; \quad (C10)$$

$$\boldsymbol{L}_1^b = i \sum_{k,\ell,p} \left( \frac{\hbar k}{2c\varepsilon_0 A_{\ell,p}^2 V} \right)^{1/2} \left[ \frac{i}{k} \hat{\boldsymbol{z}} \left\{ \sin\phi \frac{\partial}{\partial r} + \frac{i\ell}{r} \cos\phi \right\} a_{\ell,p}(k\hat{z}) f \, e^{i\ell\phi} \, e^{ikz} - H.c. \right]; \tag{C11}$$

$$\boldsymbol{T}_2^b = i \sum_{k,\ell,p} \left( \frac{\hbar k}{2c\varepsilon_0 A_{\ell,p}^2 V} \right)^{1/2} \left[ \frac{1}{k^2} \left( \hat{\boldsymbol{x}} \left\{ \sin\phi\cos\phi \frac{\partial}{\partial r}\left(\frac{\partial}{\partial r}\right) + \frac{i\ell}{r}\cos^2\phi \frac{\partial}{\partial r} - \frac{1}{r}\sin\phi\cos\phi \frac{\partial}{\partial r} \right. \right. \right.$$
$$\left. -\sin^2\phi \left(\frac{\partial}{\partial r}\frac{i\ell}{r}\right) + \frac{\ell^2}{r^2}\sin\phi\cos\phi - \frac{i\ell}{r^2}\cos^2\phi \right\} + \hat{\boldsymbol{y}} \left\{ -\cos^2\phi \frac{\partial}{\partial r}\left(\frac{\partial}{\partial r}\right) + \frac{i\ell}{r}\cos\phi\sin\phi \frac{\partial}{\partial r} - \frac{1}{r}\sin^2\phi \frac{\partial}{\partial r} \right.$$
$$\left. \left. \left. + \cos\phi\sin\phi \left(\frac{\partial}{\partial r}\frac{i\ell}{r}\right) + \frac{\ell^2}{r^2}\sin^2\phi - \frac{i\ell}{r^2}\cos\phi\sin\phi \right\} \right) a_{\ell,p}(k\hat{z}) f \, e^{i\ell\phi} \, e^{ikz} - H.c. \right]. \tag{C12}$$